\documentstyle[aps,twocolumn]{revtex}
\tightenlines
\begin{document}
\draft

\title{Thermal conductivity of amorphous carbon thin films}

\author{Andrew J. Bullen, Keith E. O'Hara, and David G. Cahill}
\address{
 Department of Materials Science and Engineering,
 Coordinated Science Laboratory, and Materials Research Laboratory,
 University of Illinois, Urbana, Illinois 61801
 }
\author{Othon Monteiro}
\address{Lawrence Berkeley Laboratory, University of California,
Berkeley, CA 94720}

\author{Achim von Keudell}
\address{Max-Planck-Intitut f\"ur Plasmaphysik, D-85740 Garching bei
M\"unchen, Germany}

\date{\today}
\maketitle
\begin{abstract}

Thermal conductivities $\Lambda $ of amorphous carbon thin films are
measured in the temperatures range 80--400~K using the $3\omega $
method.  Sample films range from soft a-C:H prepared by
remote-plasma deposition ($\Lambda = 0.20$ W m$^{-1}$ K$^{-1}$ at room
temperature) to amorphous diamond with a large fraction of $sp^3$ bonded
carbon deposited from a filtered-arc source ($\Lambda = 2.2$ W m$^{-1}$
K$^{-1}$).  Effective-medium theory provides a phenomenological
description of the variation of conductivity with mass density.  The
thermal conductivities are in good agreement with the minimum thermal
conductivity calculated from the measured atomic density and
longitudinal speed of sound.

\end{abstract}

\newpage
%\twocolumn

\section{Introduction}

Amorphous carbon (a-C) exists in an amazing variety of forms with
microstructures and physical properties that depend sensitively on
preparation method \cite{robertson}.  Because a-C thin films are often
used as protective coatings, the most thoroughly studied of these
structure-property relationships are the dependence of the mechanical
properties, e.g., elastic constants and hardness, on deposition
conditions, atomic density, and hydrogen content.  The focus of our
experimental study, thermal conductivity, like the mechanical
properties, derives from the bonding and geometry of the atomic lattice.
The large variability of microstructures within this single class of
materials provides a unique opportunity for exploring heat transport in
disordered solids \cite{orbach,feldman} and the applicability of the
minimum thermal conductivity \cite{slackrev,xtal} to materials with
heterogeneous microstructures that are common in thin films
\cite{oxide,morathdlc}.  But we also anticipate that these new data will
provide valuable insights on the high and low conductivities that can be
produced in thin film a-C for applications in the thermal engineering of
micro-devices \cite{goodsonmicro,goodsonannrev}.

The concept of a ``minimum thermal conductivity'' $\Lambda_{\rm min}$ is
based on a theory of heat transport originally proposed by Einstein
\cite{einstein}:  the atomic vibrations are assumed to be incoherent and
therefore heat diffuses between the Einstein oscillators on a time scale
of 1/2 the period of vibration.  Einstein's theory could not explain the
large thermal conductivities of most crystalline dielectrics but his
and related models \cite{slackrev,xtal,kittel} are useful for
understanding the thermal conductivity of amorphous materials and
crystals with certain types of strong disorder.

We include larger oscillating entities than the single atoms considered
by Einstein by borrowing from the Debye model of lattice vibrations and
dividing the sample into regions of size $\lambda /2$, where $\lambda$
is an acoustic wavelength, and whose frequencies
of oscillation are given by the low frequency speed of sound $\omega =
2\pi v/ \lambda$ \cite{xtal,morathdlc}.
\begin{equation}
\Lambda_{\rm min} = \left (\pi\over 6\right
)^{1/3}k_Bn^{2/3}\sum_{i=1}^3 v_i\left
(T\over \Theta _i\right )^2\int_0^{\Theta _i/T}{x^3e^x\over (e^x-1)^2}dx~~.
\end{equation}
The index $i$ labels the three sound modes (two
transverse and
one longitudinal) with speeds of sound $v_i$; $\Theta _i$ is the cutoff
frequency for each polarization expressed in degrees K, $\Theta _i =
v_i(\hbar /k_B)(6\pi ^2n)^{1/3}$, and $n$ is the number density of
atoms.  This model has no free parameters and is in good agreement with
data for a wide variety of bulk disordered materials near room
temperature \cite{xtal}.

Since diamond has the largest values of $n$ and $v_i$ \cite{mcskimin} of
any material, the high temperature limit of $\Lambda_{\rm min}$ also has
the largest possible value.  Figure 1 shows the calculated $\Lambda_{\rm
min}$ for diamond with comparisons to previously published data for
amorphous carbon \cite{morathdlc,morelli,hurler,smontara}.  Data for
bulk samples of high-dose neutron-irradiated diamond \cite{morelli} and
disordered carbon produced by high-pressure conversion of $C_{60}$
\cite{smontara} were measured by traditional steady-state methods; the
conductivities of thin film samples were measured using the mirage
effect \cite{hurler} and picosecond thermoreflectance \cite{morathdlc}.
The thin film data were measured only for room temperature, and
therefore the unusual temperature of the two bulk samples cannot be
confirmed in the thin film samples.  Furthermore, while picosecond
reflectance is a powerful probe of elastic properties and interfacial
transport of acoustic and thermal energy, measurements of thermal
conductivity using this method are relatively indirect and require
assumptions about the heat capacity of the films \cite{morathdlc}.

\section{Experimental Details}

Thin film samples of a-C:H were prepared at the Max-Planck-Institut
f\"ur Plasmaphysik by remote-plasma chemical vapor deposition
(RPCVD)---chosen to produce a soft, low-density form of a-C:H---and by
plasma-assisted CVD (PACVD); the PACVD samples have mechanical
properties that are typical for protective coatings of
``diamond-like-carbon'' (DLC) \cite{plasmadlc}.  Carbon-to-hydrogen
ratios measured on similar samples are 1:1 for RPCVD and 2:1 for PACVD.
Additional samples of DLC films were obtained from Delphi Automotive
Systems and Surmet Corporation.  At Lawrence Berkeley Laboratory, a-C
films were deposited by filtered-arc deposition (FAD)
\cite{chabot,lbldlc,iondlc} using two acceleration voltages, 100 and
2000 V. The fractions of $sp^3$ bonded carbon measured by EELS
\cite{lbldlc} on similar samples are 80\% at 100 V bias and 30\% at 2000
V bias; a-C films with low concentrations of hydrogen and carbon bonding
dominated by $sp^3$ hybridization are often referred to as ``amorphous
diamond'' (a-D) or ``tetrahedrally-bonded'' amorphous carbon (ta-C).

We use the $3\omega $ method \cite{rsi3w,cvd} to measure the thermal
conductivity of a-C films in the temperature range $80<T<400$~K.  A
$10~\mu$m wide Al line---sputter deposited on the surface of the sample
and patterned by photolithography---serves as both the heater and the
thermometer in the measurement.  If the film thickness $h$ is small
compared to the width of the metal line, heat flow is
one-dimensional in the thin film and two-dimensional (radial) in the
substrate \cite{cvd}.  Also, as long as $h$ is small compared to the
penetration depth of the thermal waves, the thin film simply adds a
frequency-independent temperature oscillation to the known thermal
response of the substrate.  Most of our a-C samples were deposited on Si
substrates with a 100 nm thick layer of thermally grown SiO$_2$
which is needed to improve the electrical isolation between the Si
substrates and the Al metallization.  The added thermal resistance of
the SiO$_2$ layer is measured separately and subtracted from the raw
data \cite{cvd}.

Conversion of the measured thermal resistance to thermal conductivity
requires accurate measurements of film thickness $h$.  We measure $h$
using spectroscopic variable-angle ellipsometry; the optical properties
of the a-C films are modeled using a fit to the resonant frequency,
oscillator strength, and damping of two Lorentz-oscillators.
Alternatively, e.g., if the optical modeling produced a poor fit to the
ellipsometry data, we use scanning electron microscopy of a fracture
cross-section to measure $h$.  Areal densities of carbon are measured
using Rutherford backscattering spectrometry of the stopping power of
the a-C film.  The combination of areal density and $h$ gives the film
density, see Table I. We measure longitudinal speeds of sound $v_l$ by
``picosecond ultrasonics'' \cite{morathdlc}:  an Al thin-film transducer
produces and detects acoustic waves generated by a mode-locked
Ti:sapphire laser operating at 780 nm; values for $v_l$ are listed in
Table I.

\section{Results and Discussion}

Figure 2 shows the results of our thermal conductivity measurements.  In
all cases, the thermal conductivity has the temperature dependence
expected for an amorphous solid in this temperature range \cite{xtal}.
In four cases, we measured the same type of film for two values of the
thickness $h$ to determine the effects of the finite thermal conductance
of interfaces on our measurements \cite{cvd}.  For the relatively low
conductivities of the RPCVD and PACVD films, see Fig.~2a, the interface
effects have little effect on the measured conductivity of films
with $h\sim 100$ nm.  Interface effects are more pronounced in the FAD
films, see Fig.~2b.  Using the assumption that the true conductivity of
the film is independent of film thickness, we can separate the true
conductivity of the film from the interface thermal conductance
\cite{cvd}; for both sets of FAD films shown in Fig.~2b, the true
conductivity is $\approx 15$\% larger than the measured conductivity of
the thicker film.

We have discovered that effective medium theory \cite{landauer} provides
a surprisingly good description of the variation of conductivity with
mass density \cite{smith}.  The conductivity of a composite structure
made of a matrix material and spherical inclusions of a second phase is
given by \cite{landauer}:
\begin{equation}
4\Lambda = (3f_2-1)\Lambda_2+(3f_1-1)\Lambda_1%
+\left[\left((3f_2-1)\Lambda_2+(3f_1-1)\Lambda_1\right)^2%
+8\Lambda_1\Lambda_2 \right]^{1/2}~ ,
\end{equation}
where $\Lambda_1$ is the conductivity of the matrix, $\Lambda_2$ is the
conductivity of the second phase, and $f_1,f_2$ are the volume fractions
of the matrix and second phase, respectively.  Figure 3 compares the
predictions of this theory to the room temperature conductivity of a-C
films.  The theory fits the data reasonably well and enables us to
extrapolate the conductivity to the full density of diamond, $\Lambda =
4.0$ W m$^{-1}$ K$^{-1}$.

Experiments on a-C have often been interpreted in terms of heterogeneous
microstructures \cite{robertsonprb,jager}, but the accuracy and
generality of these various microstructural models remains
controversial.  Our two-component model, see Eq.~2 and Fig.~3, for the
thermal conductivity is probably an oversimplification of the true
complexity of a-C microstructures.  Nevertheless, we believe this
phenomenological model will be a useful engineering guide for predicting
the conductivity of a-C films when only the density is known.

Figure 2 also includes comparisons of the data to the calculated
$\Lambda_{\rm min}$ for samples A (DLC prepared by PACVD) and H (``amorphous
diamond'' prepared by FAD).  The temperature dependence of
$\Lambda_{\rm min}$ is steeper than the data; this result is generally
observed for both bulk \cite{xtal} and thin film \cite{oxide} amorphous
materials since Eq.~1 does not include contributions to the heat
transport by low frequency phonons with long mean-free-paths
\cite{feldman}.  But in the high temperature limit, the agreement is
good, particularly for sample H, see Fig.~2b.

The agreement between the measured and calculated conductivities is made
more explicit in Fig.~4 where we compare the high temperature limit of
$\Lambda_{\rm min}$ to the data at 400 K, the highest temperature of our
measurements.  (Data for sample L are restricted to $T<300$ K because of
stray electrical conductance at higher temperatures.  In this case, we
have extrapolated the data to 400 K using the temperature dependence of
sample H.)  The calculations reproduce the trend in the data well; we
note, however, that the calculated conductivities are consistently
greater than the measured values.  The fact that the thermal
conductivities are increasing with temperature contributes to this
discrepancy; measurements at higher temperatures would show better
agreement with the model.  For the lowest conductivity films, however,
the temperature dependence of the data is relatively weak and the
calculated conductivity exceeds the measured value at 400 K by a factor
of $\approx 2$.  This relatively large difference between measured and
calculated conductivity is also observed in amorphous Se \cite{xtal}.
But given the simplifying assumptions of the model \cite{xtal},
disagreements of this magnitude are expected and we conclude that the
minimum thermal conductivity calculated from the mean atomic densities
and speeds of sound provides an adequate description of heat transport
in a wide variety of a-C thin film materials.

\section{Acknowledgements}

We thank Don Morelli and the Surmet Corp.\ for providing additional
samples of DLC.  This work was supported by NSF Grant No.\ CTS 99-78822.
Sample characterization by RBS, SEM, ellipsometry, and picosecond
ultrasonics used the facilities of the Center for Microanalysis of
Materials, supported by U.S.\ DOE Grant No.\ DEFG02-96-ER45439.

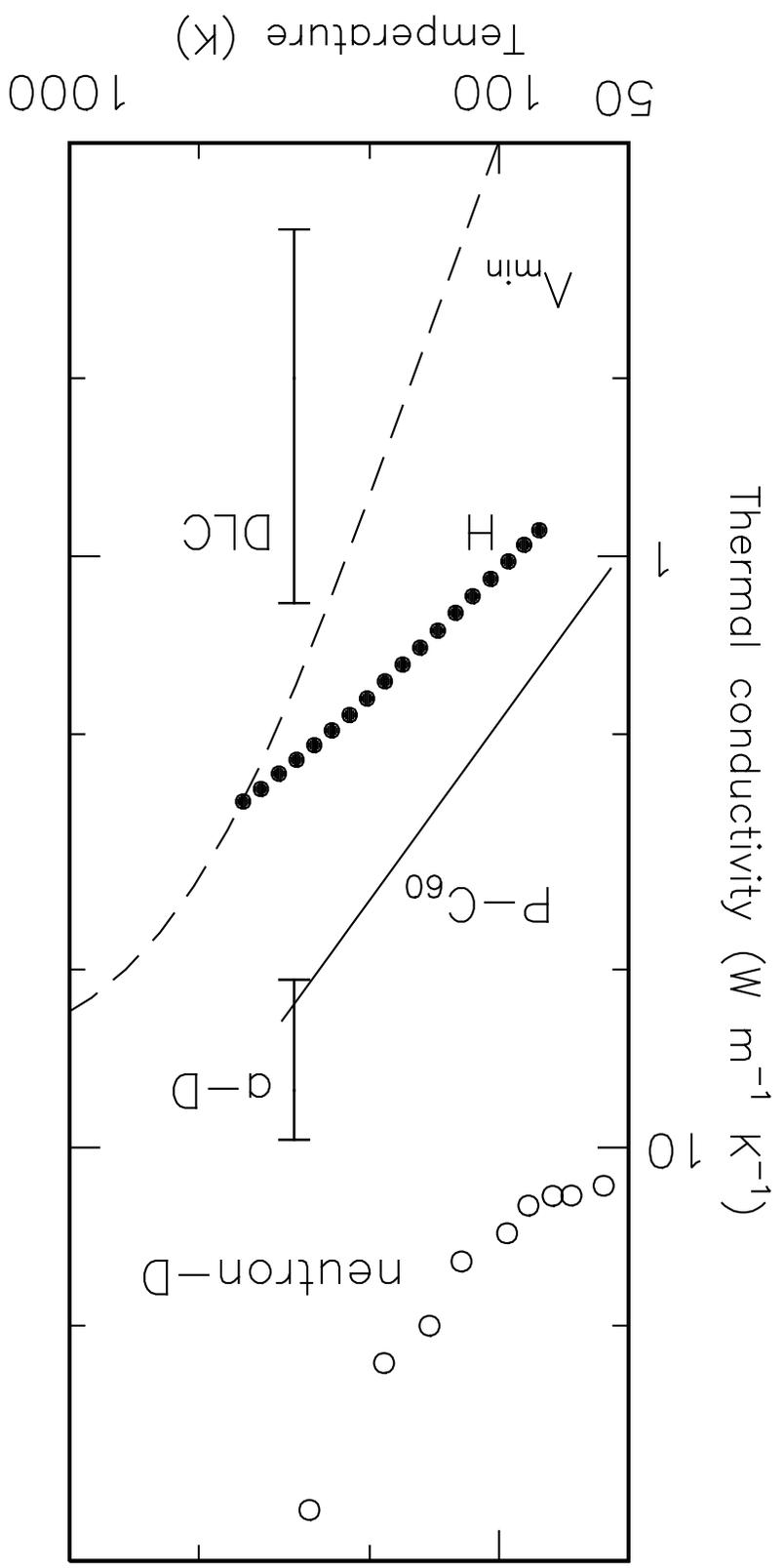
\begin{figure}

\caption{Summary of published data for amorphous carbon with comparison
to the minimum thermal conductivity $\Lambda_{\rm min}$ calculated for
diamond (dashed line) and the highest thermal conductivity measured in
this work (sample H, filled circles, see description in Table I); i)
neutron-D, diamond irradiated by neutrons at $2\times10^{22}$ cm$^{-2}$
Ref.~\protect\onlinecite{morelli}; ii) P-C$_{60}$, high-pressure
conversion of C$_{60}$, Ref.~\protect\onlinecite{smontara}; iii) DLC,
range of conductivities measured in diamond-like carbon thin films at
room temperature, Refs.~\protect\onlinecite{morathdlc} and
\protect\onlinecite{hurler}; iv) a-D, range of conductivities measured
in amorphous diamond films at room temperature,
Ref.~\protect\onlinecite{morathdlc}.}

\end{figure}

\begin{figure}

\caption{Thermal conductivity of a-C films deposited by (a)
plasma-assisted chemical vapor deposition, and (b) filtered-arc
deposition.  Data symbols are labelled by a letter that identifies each
sample; see Table I for sample descriptions; $\Lambda_{\rm min}$ for
samples A and H are shown as dashed lines in (a) and (b), respectively.}

\end{figure}

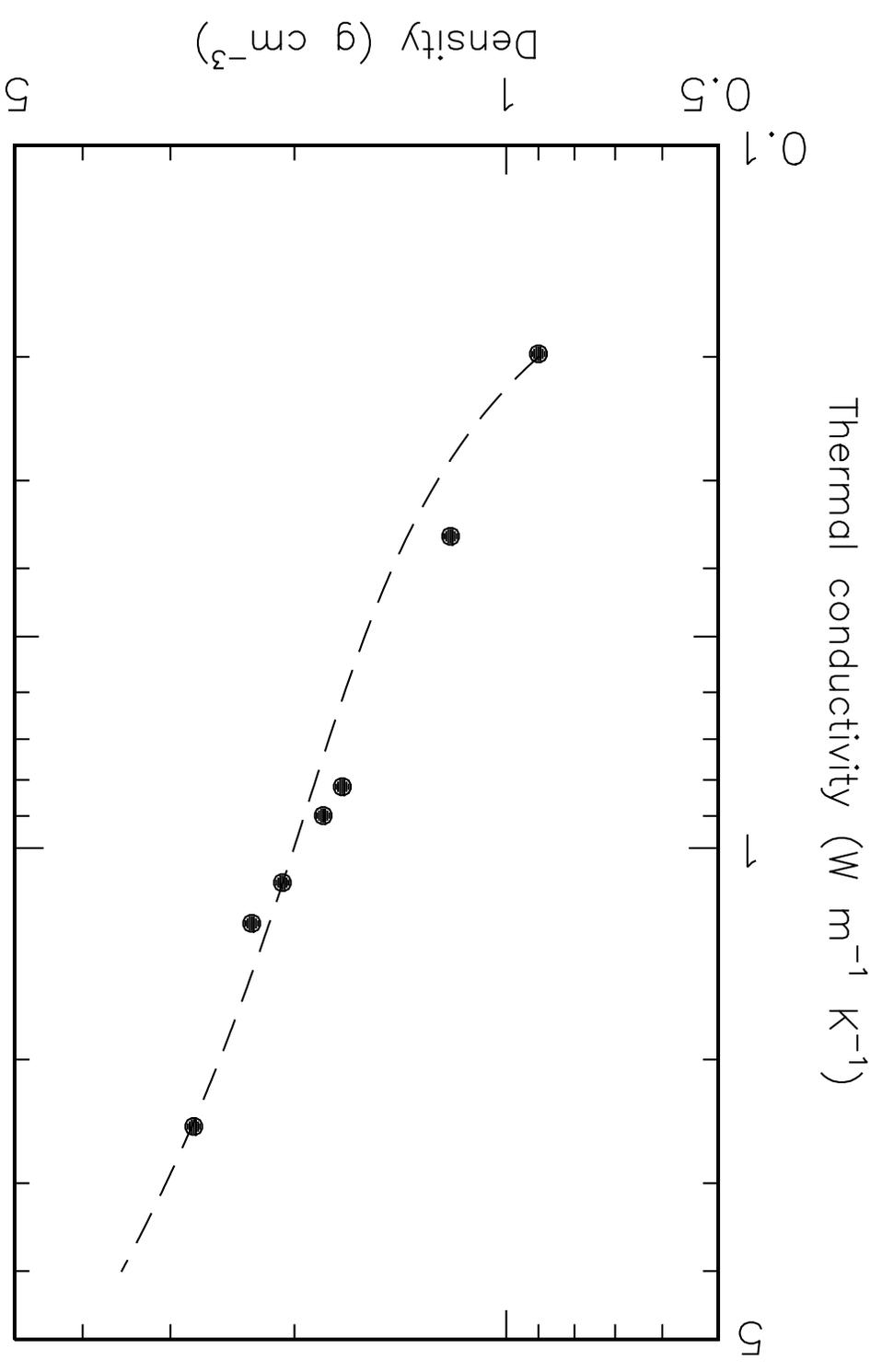
\begin{figure}

\caption{Room temperature thermal conductivity of a-C thin films as a
function of mass density.  The thermal conductivity of the filtered-arc
deposited a-C has been adjusted by a small factor ($\approx 15$\%) to
correct for the finite interface conductance.  The dashed-line is an
effective medium calculation (Eq.~2) using two-components; component 1
has the conductivity and density of our lowest conductivity film and
component 2 has the density of diamond $\rho = 3.51$ g cm$^{-3}$.  The
conductivity of component 2 is adjusted to fit the data:  $\Lambda_2 =
4.0$ W m$^{-1}$ K$^{-1}$.}

\end{figure}

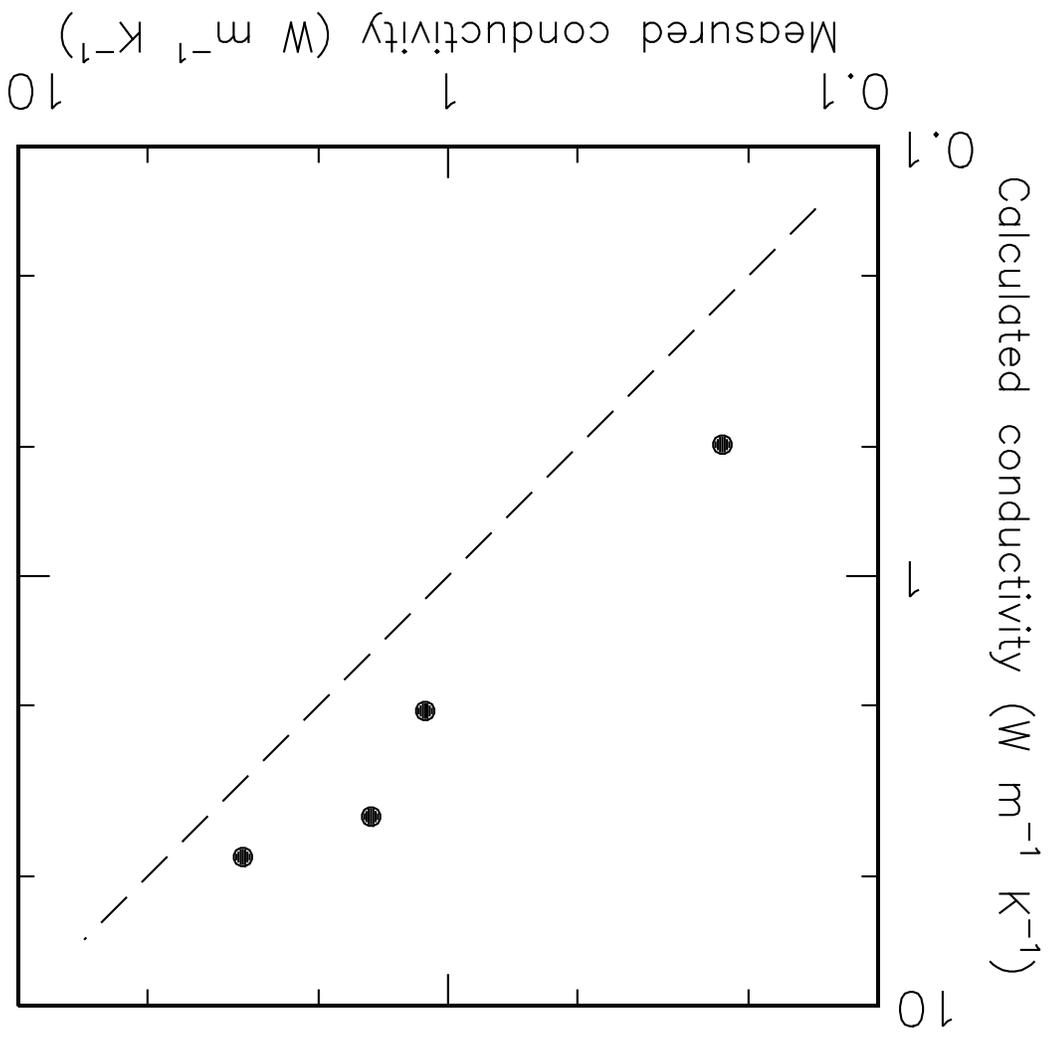
\begin{figure}

\caption{Comparison of the thermal conductivity at 400~K for samples C,
A, L, and H and the high temperature limit of Eq.~1; $\Lambda_{\rm
min}=0.40k_Bn^{2/3}(v_l+2v_t)$.  Longitudinal speeds of sounds $v_l$ are
measured by picosecond ultrasonics (see Table I); we assume $v_t \approx
0.60 v_l$ (corresponding to a Poisson's ratio of 0.22) and ignore the
contribution of hydrogen to the atomic densities $n$.  The dashed line
indicates perfect agreement between measured and calculated values.}

\end{figure}

\begin{table}
\caption{Deposition parameters and physical properties of a-C
films.  Films are deposited by plasma-assisted chemical vapor
deposition (PACVD), remote-plasma CVD (RPCVD), and filtered-arc
deposition (FAD).  Films were deposited at the Max-Planck-Institut
f\"ur Plasmaphysik (A,B,C,D), and Lawrence Berkeley Laboratory (H,I,L,M).
Additional samples were obtained from the Surmet Corporation (F,G) and Delphi
Automotive Systems (E).}

\begin{tabular}{ccccccc}
%\hline
Sample    &Film thickness  &Density           &$v_l$         &Method  & Bias \\
          &(nm)            &$\rm (g~cm^{-3})$ & km sec$^{-1}$ &       &  (volt) \\
\hline
A,B      &94,313           &1.8               & 8.7  &PACVD    &200  \\
C,D      &108,325          &0.9               & 3.4  &RPCVD    &15   \\
E        &3800             &2.1               & ---  &PACVD    &450  \\
F        &120              &1.2               & ---  &PACVD    &0    \\
G        &280              &1.7               & ---  &PACVD    &0    \\
H,I      &475,92           &2.8               & 14.0 &FAD      &100  \\
L,M      &195,65           &2.3               & 12.7 &FAD      &2000 \\
%\hline
\end{tabular}
\end{table}

\begin{references}

\bibitem{robertson} J. Robertson, Advances in Physics {\bf 35}, 317
(1986); J. Robertson, Progress in Solid State Chemistry {\bf 21}, 199
(1991).

\bibitem{orbach} R. Orbach, Philos.~Mag.  B {\bf 65}, 289 (1992).

\bibitem{feldman} J. L. Feldman, Philip B. Allen, and Scott R. Bickham,
Phys.  Rev.  B {\bf 59}, 3551 (1999).

\bibitem{slackrev} Glen A. Slack, in {\sl Solid State Physics: Advances
in Research and Applications,} edited by F. Seitz and A. G. Turnball
(Academic, New York, 1979), Vol. 34, p. 1.

\bibitem{xtal} David G. Cahill, S. K. Watson, and R. O. Pohl,
Phys.~Rev.~B {\bf 46}, 6131 (1992).

\bibitem{oxide} S.-M.~Lee, David G. Cahill, and Thomas H. Allen,
Phys.~Rev.~B {\bf 52}, 253 (1995).

\bibitem{morathdlc} C. J. Morath, H. J. Maris, J. J. Cuomo, D. L.
Pappas, A. Grill, V. V. Patel, J. P. Doyle, and K. L. Saenger, J. Appl.
Phys.  {\bf 76}, 2636 (1994).

\bibitem{goodsonmicro} K. E. Goodson, Y.-S.  Ju, and M. Asheghi, in {\sl
Microscale Energy Transport,} edited by C. L. Tien, A. Majumdar, and F.
M. Gerner, (Taylor \& Francis, New York, 1998), pp.  229-293.

\bibitem{goodsonannrev} Kenneth E. Goodson and Y. Sugtaek Ju,
Ann.\ Rev.\ Mat.\ Sci.\ {\bf 29}, 261 (1999).

\bibitem{einstein} A. Einstein, Ann.  Phys.  {\bf 35}, 679 (1911).

\bibitem{kittel} C. Kittel, Phys. Rev. {\bf 75}, 972 (1948).

\bibitem{mcskimin} H. J. McSkimin and W. L. Bond, Phys. Rev. {\bf 105},
116 (1957).

\bibitem{morelli} D. T. Morelli, T. A. Perry, J. W. Vandersande, and C.
Uher, Phys. Rev. B {\bf 48}, 3037 (1993).

\bibitem{hurler} W. Hurler, M. Pietralla, A. Hammerschmidt, Diamond \&
Related Materials {\bf 4}, 954 (1995).

\bibitem{smontara} A. Smontara, K. Biljakovic, D. Staresinic, D. Pajic,
M. E. Kozlov, M. Hirabayashi, M. Tokumoto, H. Ihara, Physica B {\bf
219\&220}, 160 (1996).

\bibitem{plasmadlc} J. W. Zou, K. Reichelt, K. Schmidt, and B. Dischler,
J. Appl.  Phys.  {\bf 65}, 3914 (1989).

\bibitem{chabot} S. Aisenberg and R. Chabot, J. Appl.  Phys.  {\bf 42},
2953 (1971).

\bibitem{lbldlc} G. M. Pharr, D. L. Callahan, S. D. McAdams, T.-Y.
Tsui, S. Anders, A. Anders, J. W. Ager III, I. G. Brown, C. S. Bhatia,
S. R. P. Silva, and J. Robertson, Appl.  Phys.  Lett.  {\bf 68}, 779
(1996).

\bibitem{iondlc} P. J. Fallon, V. S. Veerasamy, C. A. Davis, J.
Robertson, G. A. J. Amaratunga, W. I. Milne, and J. Koskinen, Phys.
Rev.  B {\bf 48} 4777 (1993).

\bibitem{rsi3w} David G. Cahill, Rev.~Sci.~Instrum. {\bf 61}, 802
(1990).

\bibitem{cvd} S.-M.~Lee and David G. Cahill, J. Appl. Phys. {\bf 81},
2590 (1997).

\bibitem{smith} Effective medium theories have often been used
for modeling the optical properties of a-C; see, for an early example,
F. W. Smith, J. Appl.  Phys.  {\bf 55}, 764 (1984).

\bibitem{robertsonprb} J. Robertson and E. P. O'Reilly, Phys.  Rev.  B
{\bf 35}, 2946 (1987).

\bibitem{jager} C. J\"ager, J. Gottwald, H. W. Spiess, and R. J. Newport,
Phys. Rev. B {\bf 50}, 846 (1994).

\bibitem{landauer} R. Landauer, J. Appl. Phys. {\bf 23}, 779 (1952).

\end{references}
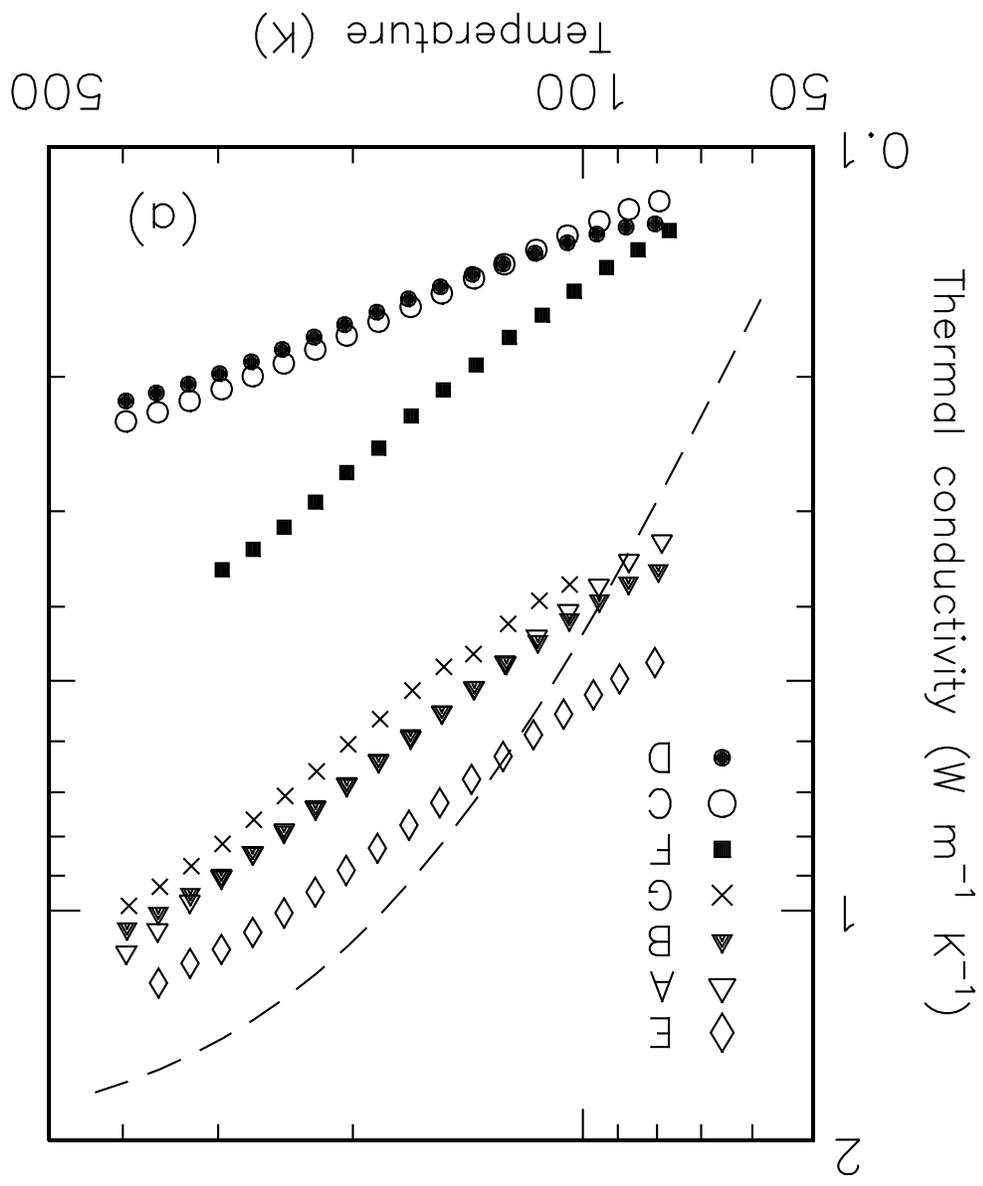
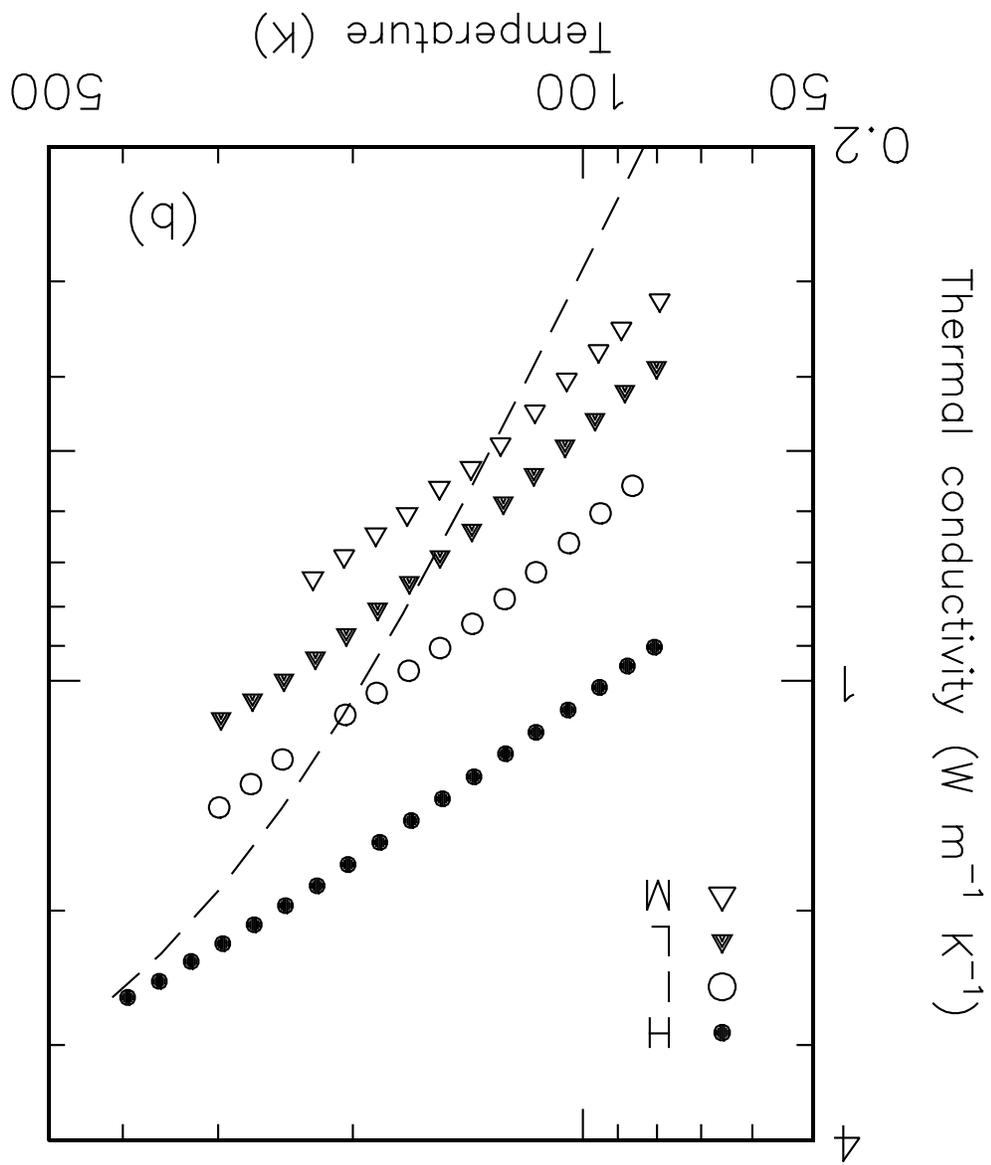
\end{document}